\documentclass[fleqn,usenatbib]{mnras}

\usepackage{newtxtext,newtxmath}

\usepackage[T1]{fontenc}
\usepackage{ae,aecompl}

\usepackage{graphicx}	
\usepackage{amsmath}	
\usepackage{amssymb}	

\newcommand{\dm}{\mathrm{DM}}
\newcommand{\fov}{{\textsc{fov}}} 

\title[Revisiting the expected \textit{Micro-X} signal from the 3.5~keV line]{Revisiting the expected \textit{Micro-X} signal from the 3.5~keV line}

\author[D.~Savchenko \& D.~Iakubovskyi]{
D.~Savchenko,$^{1}$\thanks{E-mail: dsavchenko@bitp.kiev.ua} D.~Iakubovskyi$^{1}$
\\
$^{1}$Bogolyubov Institute for Theoretical Physics, Metrolohichna str. 14b, Kyiv, Ukraine
}

\date{Accepted XXX. Received YYY; in original form ZZZ}

\pubyear{2019}

\begin{document}
\label{firstpage}
\pagerange{\pageref{firstpage}--\pageref{lastpage}}
\maketitle

\begin{abstract}
One of the future instruments to resolve the origin of the unidentified 3.5~keV emission line is the \textit{Micro-X} sounding rocket telescope. According to the estimate made in 2015, \textit{Micro-X} will be able to detect on average about 18.2~photons from the 3.5~keV line during its 300-second-long planned observation. However, this estimate is based on the extrapolation of the 3.5~keV line signal from the innermost Galactic Centre (GC) region available in 2015\@. With newly available reports on the 3.5~keV line emission in five off-centre regions, we found that similar \textit{Micro-X} payload will result in 3.4--4.3 counts on average, depending on the dark matter distribution. Therefore, we show that the 3.5~keV line is unlikely to be detected with a single \textit{Micro-X} launch using an \emph{original} \textit{Micro-X} payload. Increasing its field-of-view from 20$^\circ$ to 33$^\circ$ and its repointing out of GC (to avoid the brightest X-ray point source on the sky, Sco X-1) will increase the expected number of counts from 3.5~keV line to 7.5--7.9, which corresponds to its expected marginal ($\sim 2\sigma$) detection within a single \textit{Micro-X} observation.
\end{abstract}




\section{Introduction}

The origin of the unidentified 3.5~keV emission line remains debated, see~\cite{Adhikari:2016bei, Abazajian:2017tcc, Boyarsky:2018tvu} for reviews. So far, this line is reported from a variety of cosmic objects, such as the combined spectrum of bright galaxy clusters and the central part of Perseus cluster~\citep{Bulbul:2014sua}, the central part of Andromeda galaxy and the outskirts of Perseus cluster~\citep{Boyarsky:2014jta}, the innermost part of our Galaxy~\citep{Riemer-Sorensen:2014yda, Jeltema:2014qfa, Boyarsky:2014ska}, a number of nearby individual galaxy clusters~\citep{Urban:2014yda, Iakubovskyi:2015dna}, and the parts of Galactic halo far away from the Galactic Center~\citep{Neronov:2016wdd, Cappelluti:2017ywp} and close to the Galactic Center~\citep{Boyarsky:2018ktr, Hofmann:2019ihc} (see, however, the results of~\citet{Dessert:18}, who had not reported the presence of the 3.5~keV line in 5$^\circ$ -- 45$^\circ$ region). While initially the 3.5~keV line was interpreted as a signal from decaying dark matter~\citep{Bulbul:2014sua,Boyarsky:2014jta, Boyarsky:2014ska}, other hypotheses about the line origin, including emission by highly excited Potassium~\citep{Jeltema:2014qfa, Phillips:15} and Sulphur~\citep{Gu:2015gqm, Shah:2016efh} ions, also attracted much attention.  

Unfortunately, the moderate energy resolution (with the full width at half maximum (FWHM) about 50--100~eV) of modern CCD-based X-ray telescopes, such as \textit{XMM-Newton} and \textit{Chandra}, does not allow one to robustly distinguish between the astrophysical and dark matter scenarios. Instead, the upcoming X-ray imaging spectrometers with several-eV FWHM energy resolution were proposed to resolve the 3.5~keV origin, see, e.g., \cite{Bulbul:2014sua, Iakubovskyi:2015kwa, Speckhard:2015eva}. While the \textit{Hitomi} (former \textit{Astro-H}) satellite, because of its unexpected abrupt failure, did not collect enough data to significantly detect the 3.5~keV line~\citep{Aharonian:2016gzq}, other cosmic missions with eV-scale resolution spectrometres, such as \textit{Micro-X},\footnote{\url{https://microx.northwestern.edu/}} \textit{XRISM}\footnote{\url{https://heasarc.gsfc.nasa.gov/docs/xrism/about/}} (former \textit{Hitomi} recovery mission), and Lynx\footnote{\url{https://wwwastro.msfc.nasa.gov/lynx/}} are planned to be launched during the forthcoming several years.

\cite{Figueroa-Feliciano:2015gwa} estimate the observed 3.5~keV signal for the earliest among the planned missions --- the \textit{Micro-X} sounding rocket-based spectrometer. Specifically, they expect 18.2 counts from the planned 300~second-long observation of the circular region around the Galactic Centre with 20$^\circ$  field-of-view (FoV) radius. For comparison, the 2$\sigma$ line detection limit calculated by~\cite{Figueroa-Feliciano:2015gwa} is only about 6 counts at 3.5~keV.

However, the signal estimate by~\cite{Figueroa-Feliciano:2015gwa} is based on the \emph{extrapolation} of the 3.5~keV line intensity reported by~\cite{Boyarsky:2014ska} in a much smaller ($14'$ radius) circle around the Galactic Centre. After the work by~\cite{Boyarsky:2018ktr}, more detections out of Galactic Centre (up to 35$^\circ$ radius) became available. This allows us to improve the signal estimate by~\cite{Figueroa-Feliciano:2015gwa}, avoiding extrapolations. 

\section{Revisiting the expected 3.5~keV signal}

The total number of counts from a decaying dark matter observation is~\citep[see, e.g.,][]{Boyarsky:2007ge}
\begin{equation}
C_\dm = \frac{A_\text{eff}T_\text{exp}\Gamma_\dm}{4\pi m_\dm}\mathcal{I}, \qquad \mathcal{I} = \int\frac{V(\theta)
\rho_\dm(\textbf{r})d^3\textbf{r}}{|{\bf r}|^2},\label{eq:dark-matter-decay-rate}
\end{equation}
where $T_\text{exp}$ is the observation time, $\Gamma_\dm$ is the radiative dark matter decay width, $m_\dm$ is the mass of dark matter particle,
$\rho_\dm$ is the dark matter mass density distribution, $\textbf{r}$ is the radius-vector calculated from 
the \emph{observer's} position,
$A_\text{eff}$ is the effective area of the \textit{Micro-X} micro-calorimeter for an off-axis source, $\theta$ is the angle between $\textbf{r}$ and the field-of-view (FoV) axis, and $V(\theta) = A_\text{eff}(\mathbf{r})/A_\text{eff}$ is the vignetting correction factor. 

Similar to~\cite{Figueroa-Feliciano:2015gwa}, we choose $V(\theta) = 1$. Then the integral $\mathcal{I}$ in Eq.~(\ref{eq:dark-matter-decay-rate}) is equal to
 \begin{equation}
 2\pi \int_0^{\theta_\fov} \sin\theta d\theta 
   \int_0^{z_0(\theta)} dz \rho_\dm\left(\sqrt{r_\odot^2+z^2-2zr_\odot\cos\theta}\right),
   \end{equation}
where  $z_0(\theta) = r_\odot\cos\theta + \sqrt{R_\text{max}^2 - r_\odot^2\sin^2\theta}$,
$\theta_\fov$ is the FoV radius (in radians), $r_\odot$ is the distance to the Galactic Centre, 
$R_\text{max}$ is the maximal radius of the dark matter halo (which we assumed to coincide with the halo virial radius). As a reference value, we take $r_\odot = 8.127\pm 0.031$~kpc\footnote{Varying $r_\odot$ by $\sim$0.1~kpc, one changes the expected \textit{Micro-X} numbers of counts change by $\sim$0.6~cts ($\sim$15\%).} from~\cite{Abuter:2018drb}.

Similar to~\cite{Boyarsky:2018ktr}, we used three types of dark matter distributions:

\begin{enumerate}
    \item Navarro--Frenk--White (NFW) profile~\citep{Navarro:1995iw, Navarro:1996gj};
    \item Burkert (BURK) profile~\citep{Burkert:1995yz};
    \item Einasto (EIN) profile~\citep{Einasto:65}.
\end{enumerate}
For each of these profiles, we use the characteristic parameters and values of $\Gamma_\dm$ that correspond to their best-fit of 3.5~keV line fluxes reported in five concentric circles with $10'$--$14'$, $14'$--3$^\circ$, 3$^\circ$--10$^\circ$, 10$^\circ$--20$^\circ$ and 20$^\circ$--35$^\circ$, see Table~II of \cite{Boyarsky:2018ktr} for details. We also use the same planned observation duration (300~s) and effective area (1~cm$^2$) of a micro-calorimeter on-board \textit{Micro-X} as in the original paper by~\cite{Figueroa-Feliciano:2015gwa}.

\section{Results and conclusions}

First, we assumed the same radius of \textit{Micro-X} FoV (20$^\circ$ circle), pointed towards the Galactic Centre, as in~\cite{Figueroa-Feliciano:2015gwa}. However, the calculated expected number of counts from the 3.5~keV line is much smaller compared to the result of \cite{Figueroa-Feliciano:2015gwa} (18.2~counts). Specifically, for our NFW profile, we expect only 3.8~counts for a single \textit{Micro-X} observation. The expected numbers for BURK and EIN profiles are 3.4 and 4.3~counts, respectively.

\begin{figure}
    \centering
    \includegraphics[width=\columnwidth]{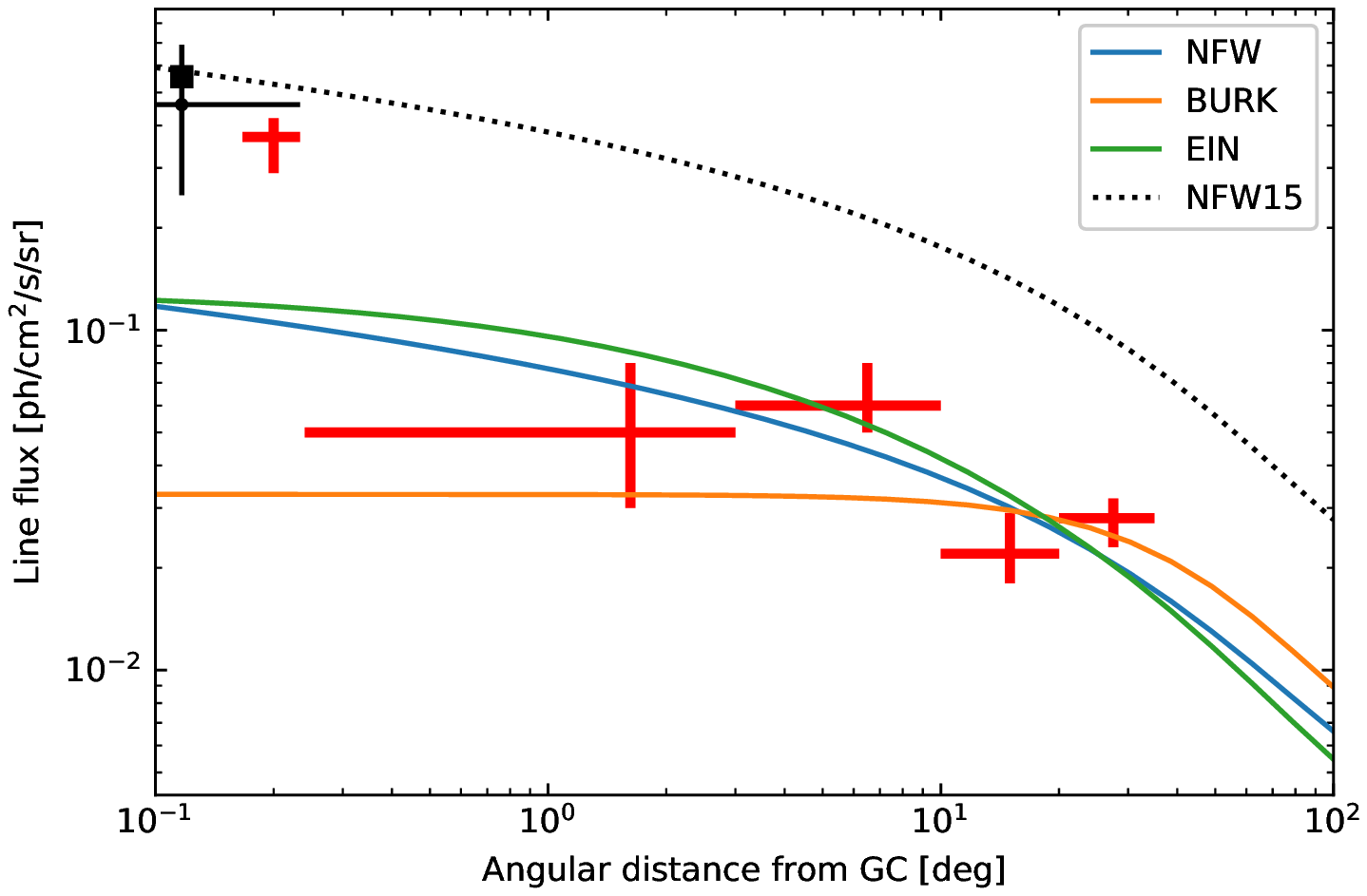}
    \caption{Expected 3.5~keV line fluxes (as functions of the angular distance from the Galactic Centre) obtained for different dark matter distributions, including NFW, EIN and BURK profiles used in this paper, and the NFW15 profile used in \protect\cite{Figueroa-Feliciano:2015gwa}. 
    The red crosses show datapoints from~\protect\citet{Boyarsky:2018ktr}. 
    The black square corresponds to the reference flux $29\times10^{-6}$\,cts/sec/cm$^2$ from~\protect\cite{Boyarsky:2014ska}, used by~\protect\cite{Figueroa-Feliciano:2015gwa} to obtain their estimates.
    The black cross corresponds to the flux $(24\pm12)\times10^{-6}$\,cts/sec/cm$^2$, also reported in~\protect\cite{Boyarsky:2014ska}, with possible contribution from the widened Ar~XVII line complex at 3.685~keV taken into account, similarly to~\protect\cite{Boyarsky:2018ktr}.}
    \label{fig:microx-brightness}
\end{figure}

The reason behind the significant decrease of the estimated number of counts compared to the work by~\cite{Figueroa-Feliciano:2015gwa} is a much steeper decline of the observed 3.5~keV line flux outside the innermost 14$'$ circle, compared to the extrapolation by~\cite{Figueroa-Feliciano:2015gwa}.
Fig.~\ref{fig:microx-brightness} of this paper illustrates this behaviour. 

All of our new 3.5~keV line count estimates are \emph{below the 2$\sigma$ detection limit} by \textit{Micro-X} of about 6~counts, see Fig.~11 from \cite{Figueroa-Feliciano:2015gwa} for details. This means that a single \textit{Micro-X} observation of the Galactic Centre region with 20$^\circ$ FoV radius is very unlikely to detect the 3.5~keV line. A straightforward approach will thus require a combination of many \textit{Micro-X} observations to detect the line. An alternative is to increase the FoV of the \textit{Micro-X}; for example, its predecessor, XQC, has about 33$^\circ$ FoV radius~\citep{Figueroa-Feliciano:2015gwa}. The problem, however, is in the presence of a very bright X-ray object --- Sco~X-1 --- located only about 24$^\circ$ from the Galactic Centre. To limit the background counts, the increase in the \textit{Micro-X} FoV will also require its repointing away from the Galactic Centre. For example, using the same dark matter distributions from~\cite{Boyarsky:2018ktr}, the expected number of 3.5~keV line counts from a 33$^\circ$ FoV located 13$^\circ$ away from the Galactic Centre (to avoid emission from Sco X-1) is 7.5--7.9. In combination with a modest increase of the number of expected background counts in larger FoV (in the absence of bright X-ray sources), such an increase of the instrument FoV with its subsequent re-pointing may lead to marginal ($\sim 2\sigma$) detection of 3.5~keV line even with a \emph{single} \textit{Micro-X} launch.

When this paper was finished, we were aware of the forthcoming paper of \texttt{Micro-X} collaboration~\citep{MicroX-new}. Remarkably, our results agree with those presented in Table~2 of~\cite{MicroX-new}. For example, for their \textit{Micro-X} North target ($l = 31^\circ$, $b = 40^\circ$), we calculated that a single \textit{Micro-X} observation with 33$^\circ$ radius circle FoV will collect 4.5~cts for NFW, 5.9~cts for BURK and 4.2~cts for EIN profile from 3.5~keV line.
For their \textit{Micro-X} South target ($l = 0^\circ$, $b = -12^\circ$) we expect 7.5--7.9 counts from a single observation.

Throughout this paper we have used the 3.5~keV line intensities, obtained by~\citet{Boyarsky:2018ktr} using \textit{XMM-Newton} data. While the reported values are consistent with recent result of~\citet{Hofmann:2019ihc} (who detected the signal in \textit{Chandra} data), there is an apparent discrepancy between the results of~\citet{Boyarsky:2018ktr} and the findings of~\citet{Dessert:18}, who have not reported the line in the \textit{XMM-Newton} observations of the 5$^\circ$ -- 45$^\circ$ region. In~\citet{Boyarsky:2018ktr} we describe in details our view on discrepancies with~\citet{Dessert:18}, which, we believe, are in choice of the underlying background model. As fig.~5 in~\citet{Boyarsky:2018ktr} demonstrates, by using narrow fitting interval of $3.3-3.8$~keV, as in~\citet{Dessert:18}, one tend to over-predict the background level, effectively ``masking'' the 3.5~keV signal. 

\section*{Acknowledgements}

The authors are grateful to Yu. Shtanov for  collaboration  and  valuable  comments. This work was supported by the grant for young scientist research laboratories of the National Academy of Sciences of Ukraine.




\bibliographystyle{mnras}
\bibliography{microx}

\begin{thebibliography}{}
\makeatletter
\relax
\def\mn@urlcharsother{\let\do\@makeother \do\$\do\&\do\#\do\^\do\_\do\%\do\~}
\def\mn@doi{\begingroup\mn@urlcharsother \@ifnextchar [ {\mn@doi@}
  {\mn@doi@[]}}
\def\mn@doi@[#1]#2{\def\@tempa{#1}\ifx\@tempa\@empty \href
  {http://dx.doi.org/#2} {doi:#2}\else \href {http://dx.doi.org/#2} {#1}\fi
  \endgroup}
\def\mn@eprint#1#2{\mn@eprint@#1:#2::\@nil}
\def\mn@eprint@arXiv#1{\href {http://arxiv.org/abs/#1} {{\tt arXiv:#1}}}
\def\mn@eprint@dblp#1{\href {http://dblp.uni-trier.de/rec/bibtex/#1.xml}
  {dblp:#1}}
\def\mn@eprint@#1:#2:#3:#4\@nil{\def\@tempa {#1}\def\@tempb {#2}\def\@tempc
  {#3}\ifx \@tempc \@empty \let \@tempc \@tempb \let \@tempb \@tempa \fi \ifx
  \@tempb \@empty \def\@tempb {arXiv}\fi \@ifundefined
  {mn@eprint@\@tempb}{\@tempb:\@tempc}{\expandafter \expandafter \csname
  mn@eprint@\@tempb\endcsname \expandafter{\@tempc}}}

\bibitem[\protect\citeauthoryear{Abazajian}{Abazajian}{2017}]{Abazajian:2017tcc}
Abazajian K.~N.,  2017, \mn@doi [Phys. Rept.] {10.1016/j.physrep.2017.10.003},
  711-712, 1

\bibitem[\protect\citeauthoryear{Abuter et~al.}{Abuter
  et~al.}{2018}]{Abuter:2018drb}
Abuter R.,  et~al., 2018, \mn@doi [Astron. Astrophys.]
  {10.1051/0004-6361/201833718}, 615, L15

\bibitem[\protect\citeauthoryear{{Adams} et~al.,}{{Adams}
  et~al.}{2019}]{MicroX-new}
{Adams} J.~S.,  et~al., 2019, preprint, \href
  {https://ui.adsabs.harvard.edu/abs/2019arXiv190809010A} {} (\mn@eprint
  {arXiv} {1908.09010})

\bibitem[\protect\citeauthoryear{Aharonian et~al.}{Aharonian
  et~al.}{2017}]{Aharonian:2016gzq}
Aharonian F.~A.,  et~al., 2017, \mn@doi [Astrophys. J.]
  {10.3847/2041-8213/aa61fa}, 837, L15

\bibitem[\protect\citeauthoryear{Boyarsky, Malyshev, Neronov  \&
  Ruchayskiy}{Boyarsky et~al.}{2008}]{Boyarsky:2007ge}
Boyarsky A.,  Malyshev D.,  Neronov A.,   Ruchayskiy O.,  2008, \mn@doi [Mon.
  Not. Roy. Astron. Soc.] {10.1111/j.1365-2966.2008.13003.x}, 387, 1345

\bibitem[\protect\citeauthoryear{Boyarsky, Ruchayskiy, Iakubovskyi  \&
  Franse}{Boyarsky et~al.}{2014}]{Boyarsky:2014jta}
Boyarsky A.,  Ruchayskiy O.,  Iakubovskyi D.,   Franse J.,  2014, \mn@doi
  [Phys. Rev. Lett.] {10.1103/PhysRevLett.113.251301}, 113, 251301

\bibitem[\protect\citeauthoryear{Boyarsky, Franse, Iakubovskyi  \&
  Ruchayskiy}{Boyarsky et~al.}{2015}]{Boyarsky:2014ska}
Boyarsky A.,  Franse J.,  Iakubovskyi D.,   Ruchayskiy O.,  2015, \mn@doi
  [Phys. Rev. Lett.] {10.1103/PhysRevLett.115.161301}, 115, 161301

\bibitem[\protect\citeauthoryear{Boyarsky, Iakubovskyi, Ruchayskiy  \&
  Savchenko}{Boyarsky et~al.}{2018}]{Boyarsky:2018ktr}
Boyarsky A.,  Iakubovskyi D.,  Ruchayskiy O.,   Savchenko D.,  2018, preprint
  (\mn@eprint {arXiv} {1812.10488})

\bibitem[\protect\citeauthoryear{Boyarsky, Drewes, Lasserre, Mertens  \&
  Ruchayskiy}{Boyarsky et~al.}{2019}]{Boyarsky:2018tvu}
Boyarsky A.,  Drewes M.,  Lasserre T.,  Mertens S.,   Ruchayskiy O.,  2019,
  \mn@doi [Prog. Part. Nucl. Phys.] {10.1016/j.ppnp.2018.07.004}, 104, 1

\bibitem[\protect\citeauthoryear{Bulbul, Markevitch, Foster, Smith, Loewenstein
   \& Randall}{Bulbul et~al.}{2014}]{Bulbul:2014sua}
Bulbul E.,  Markevitch M.,  Foster A.,  Smith R.~K.,  Loewenstein M.,   Randall
  S.~W.,  2014, \mn@doi [Astrophys. J.] {10.1088/0004-637X/789/1/13}, 789, 13

\bibitem[\protect\citeauthoryear{Burkert}{Burkert}{1996}]{Burkert:1995yz}
Burkert A.,  1996, \mn@doi [IAU Symp.] {10.1086/309560}, 171, 175

\bibitem[\protect\citeauthoryear{Cappelluti et~al.,}{Cappelluti
  et~al.}{2018}]{Cappelluti:2017ywp}
Cappelluti N.,  et~al., 2018, \mn@doi [Astrophys. J.]
  {10.3847/1538-4357/aaaa68}, 854, 179

\bibitem[\protect\citeauthoryear{{Dessert}, {Rodd}  \& {Safdi}}{{Dessert}
  et~al.}{2018}]{Dessert:18}
{Dessert} C.,  {Rodd} N.~L.,   {Safdi} B.~R.,  2018, preprint, \href
  {https://ui.adsabs.harvard.edu/abs/2018arXiv181206976D} {} (\mn@eprint
  {arXiv} {1812.06976})

\bibitem[\protect\citeauthoryear{Drewes et~al.}{Drewes
  et~al.}{2017}]{Adhikari:2016bei}
Drewes M.,  et~al., 2017, \mn@doi [JCAP] {10.1088/1475-7516/2017/01/025}, 1701,
  025

\bibitem[\protect\citeauthoryear{{Einasto}}{{Einasto}}{1965}]{Einasto:65}
{Einasto} J.,  1965, Trudy Astrofizicheskogo Instituta Alma-Ata, \href
  {https://ui.adsabs.harvard.edu/abs/1965TrAlm...5...87E} {5, 87}

\bibitem[\protect\citeauthoryear{Figueroa-Feliciano et~al.}{Figueroa-Feliciano
  et~al.}{2015}]{Figueroa-Feliciano:2015gwa}
Figueroa-Feliciano E.,  et~al., 2015, \mn@doi [Astrophys. J.]
  {10.1088/0004-637X/814/1/82}, 814, 82

\bibitem[\protect\citeauthoryear{Gu, Kaastra, Raassen, Mullen, Cumbee, Lyons
  \& Stancil}{Gu et~al.}{2015}]{Gu:2015gqm}
Gu L.,  Kaastra J.,  Raassen A. J.~J.,  Mullen P.~D.,  Cumbee R.~S.,  Lyons D.,
    Stancil P.~C.,  2015, \mn@doi [Astron. Astrophys.]
  {10.1051/0004-6361/201527634}, 584, L11

\bibitem[\protect\citeauthoryear{Hofmann \& Wegg}{Hofmann \&
  Wegg}{2019}]{Hofmann:2019ihc}
Hofmann F.,  Wegg C.,  2019, \mn@doi [Astron. Astrophys.]
  {10.1051/0004-6361/201935561}, 625, L7

\bibitem[\protect\citeauthoryear{Iakubovskyi}{Iakubovskyi}{2015}]{Iakubovskyi:2015kwa}
Iakubovskyi D.,  2015, \mn@doi [Mon. Not. Roy. Astron. Soc.]
  {10.1093/mnras/stv1955}, 453, 4097

\bibitem[\protect\citeauthoryear{Iakubovskyi, Bulbul, Foster, Savchenko  \&
  Sadova}{Iakubovskyi et~al.}{2015}]{Iakubovskyi:2015dna}
Iakubovskyi D.,  Bulbul E.,  Foster A.~R.,  Savchenko D.,   Sadova V.,  2015,
  preprint (\mn@eprint {arXiv} {1508.05186})

\bibitem[\protect\citeauthoryear{Jeltema \& Profumo}{Jeltema \&
  Profumo}{2015}]{Jeltema:2014qfa}
Jeltema T.~E.,  Profumo S.,  2015, \mn@doi [Mon. Not. Roy. Astron. Soc.]
  {10.1093/mnras/stv768}, 450, 2143

\bibitem[\protect\citeauthoryear{Navarro, Frenk  \& White}{Navarro
  et~al.}{1996}]{Navarro:1995iw}
Navarro J.~F.,  Frenk C.~S.,   White S. D.~M.,  1996, \mn@doi [Astrophys. J.]
  {10.1086/177173}, 462, 563

\bibitem[\protect\citeauthoryear{Navarro, Frenk  \& White}{Navarro
  et~al.}{1997}]{Navarro:1996gj}
Navarro J.~F.,  Frenk C.~S.,   White S. D.~M.,  1997, \mn@doi [Astrophys. J.]
  {10.1086/304888}, 490, 493

\bibitem[\protect\citeauthoryear{Neronov, Malyshev  \& Eckert}{Neronov
  et~al.}{2016}]{Neronov:2016wdd}
Neronov A.,  Malyshev D.,   Eckert D.,  2016, \mn@doi [Phys. Rev.]
  {10.1103/PhysRevD.94.123504}, D94, 123504

\bibitem[\protect\citeauthoryear{{Phillips}, {Sylwester}  \&
  {Sylwester}}{{Phillips} et~al.}{2015}]{Phillips:15}
{Phillips} K.~J.~H.,  {Sylwester} B.,   {Sylwester} J.,  2015, \mn@doi [\apj]
  {10.1088/0004-637X/809/1/50}, \href
  {https://ui.adsabs.harvard.edu/abs/2015ApJ...809...50P} {809, 50}

\bibitem[\protect\citeauthoryear{Riemer-S{\o}rensen}{Riemer-S{\o}rensen}{2016}]{Riemer-Sorensen:2014yda}
Riemer-S{\o}rensen S.,  2016, \mn@doi [Astron. Astrophys.]
  {10.1051/0004-6361/201527278}, 590, A71

\bibitem[\protect\citeauthoryear{{Shah}, {Dobrodey}, {Bernitt},
  {Steinbr{\"u}gge}, {Crespo L{\'o}pez-Urrutia}, {Gu}  \& {Kaastra}}{{Shah}
  et~al.}{2016}]{Shah:2016efh}
{Shah} C.,  {Dobrodey} S.,  {Bernitt} S.,  {Steinbr{\"u}gge} R.,  {Crespo
  L{\'o}pez-Urrutia} J.~R.,  {Gu} L.,   {Kaastra} J.,  2016, \mn@doi [\apj]
  {10.3847/1538-4357/833/1/52}, \href
  {https://ui.adsabs.harvard.edu/abs/2016ApJ...833...52S} {833, 52}

\bibitem[\protect\citeauthoryear{Speckhard, Ng, Beacom  \& Laha}{Speckhard
  et~al.}{2016}]{Speckhard:2015eva}
Speckhard E.~G.,  Ng K. C.~Y.,  Beacom J.~F.,   Laha R.,  2016, \mn@doi [Phys.
  Rev. Lett.] {10.1103/PhysRevLett.116.031301}, 116, 031301

\bibitem[\protect\citeauthoryear{Urban, Werner, Allen, Simionescu, Kaastra  \&
  Strigari}{Urban et~al.}{2015}]{Urban:2014yda}
Urban O.,  Werner N.,  Allen S.~W.,  Simionescu A.,  Kaastra J.~S.,   Strigari
  L.~E.,  2015, \mn@doi [Mon. Not. Roy. Astron. Soc.] {10.1093/mnras/stv1142},
  451, 2447

\makeatother
\end{thebibliography}






\bsp	
\label{lastpage}
\end{document}